\documentclass{article}
\usepackage{arxiv}
\usepackage[utf8]{inputenc} 
\usepackage[T1]{fontenc}    
\usepackage{hyperref}       
\usepackage{url}            
\usepackage{booktabs}       
\usepackage{amsmath,amsthm,mathtools,amssymb,enumitem}
\usepackage{nicefrac}       
\usepackage{microtype}      
\usepackage{graphicx,xcolor}
\usepackage{endnotes}

\DeclareMathOperator*{\plim}{plim}
\theoremstyle{definition}
\newtheorem{theorem}{Theorem}[section] 
\newtheorem{definition}[theorem]{Definition} 
\newtheorem{remark}[theorem]{Remark}
\newtheorem{lemma}[theorem]{Lemma}

\usepackage{listings}
\lstdefinelanguage{R}{
  keywords={if,else,repeat,while,function,for,in,next,break,TRUE,FALSE,NA,NaN,Inf},
  sensitive=true,
  comment=[l]\#,
  morestring=[b]"
}
\title{Generalized Jeffreys's approximate objective \\ Bayes factor: Model-selection consistency, \\ finite-sample accuracy, and detecting the type I errors \\ in 71,126 clinical trial findings}

\author{
 Puneet Velidi \\
  Department of Mathematics and Statistics\\
  University of Victoria\\
  Victoria, BC V8W 2Y2 \\
  \texttt{pvelidi@uvic.ca} \\
  \And
 Zhengxiao Wei \\
  Department of Mathematics and Statistics\\
  University of Victoria\\
  Victoria, BC V8W 2Y2 \\
  \texttt{zhengxiao@uvic.ca} \\
  \And
 Shreena Nisha Kalaria \\
  Deeley Research Centre\\
  British Columbia Cancer Agency\\
  Victoria, BC V8R 6V5 \\
  \texttt{skalaria@bccrc.ca} \\
  \And
 Yimeng Liu \\
  Department of Mathematics and Statistics\\
  University of Victoria\\
  Victoria, BC V8W 2Y2 \\
  \texttt{yimengliu@uvic.ca} \\
  \And
 Céline M. Laumont \\
  Deeley Research Centre\\
  British Columbia Cancer Agency\\
  Victoria, BC V8R 6V5 \\
  \texttt{claumont@bccrc.ca} \\
  \And
 Brad H. Nelson \\
  Deeley Research Centre\\
  British Columbia Cancer Agency\\
  Victoria, BC V8R 6V5 \\
  \texttt{bnelson@bccrc.ca} \\
  \And
 Farouk S. Nathoo \\
  Department of Mathematics and Statistics\\
  University of Victoria\\
  Victoria, BC V8W 2Y2 \\
  \texttt{nathoo@uvic.ca} \\
}

\begin{document}
\maketitle
\begin{abstract}
Concerns about the misuse and misinterpretation of $p$-values and statistical significance have motivated alternatives for quantifying evidence. We define a generalized form of Jeffreys's approximate objective Bayes factor ($\mathit{eJAB}$), a one-line calculation that is a function of the $p$-value, sample size, and dimension of the test. We establish conditions under which $\mathit{eJAB}$ is model-selection consistent and verify them for ten common statistical tests. We assess finite-sample accuracy by comparing $\mathit{eJAB}$ with Markov chain Monte Carlo computed Bayes factors in 12 simulation studies. We then apply $\mathit{eJAB}$ to 71,126 results from ClinicalTrials.gov (CTG) and find that the proportion of findings with $p\text{-value}\leqslant\alpha$ yet $\mathit{eJAB}_{01}>1$ is close to $\alpha$ over a wide range, suggesting that such contradictions are pointing to the type I errors. We catalog 4,088 such candidate type I errors and provide details for 131 with reported $p$-value $\leqslant .01$. We also identify 487 instances of the Jeffreys–Lindley paradox in clinical trials. Finally, we estimate that 75\% (6\%) of clinical trial plans from CTG set $\alpha\geqslant.05$ as the target evidence threshold, and that 35.5\% (0.22\%) of results significant at $\alpha=.05$ correspond to evidence that is no stronger than anecdotal under $\mathit{eJAB}$.
\end{abstract}


\bigskip
\bigskip

    Concerns about the replicability of scientific results—stemming in part from the overreliance on null-hypothesis significance testing (NHST) at traditional $p$-value thresholds \cite{Ioannidis2005-uq}—have prompted calls for more stringent evidence criteria \cite{benjamin2018redefine, johnson2013revised}, more judicious interpretation \cite{wasserstein2016asa}, and alternative measures of evidence \cite{berger1987testing, grunwald2024beyond, evans2015measuring, halsey2019reign}. The Bayes factor—the ratio of marginal likelihoods under competing hypotheses—can provide evidence for either the null or the alternative \cite{kass1995bayes, ly2016evaluation, raftery1995bayesian, morey2016philosophy, wagenmakers2007practical, kass1995reference}. It represents how data update the prior model odds to form the posterior model odds. The evidential behavior of both $p$-values and Bayes factors has been analyzed by stochastic ordering in sample size and effect size, with implications for the Jeffreys–Lindley paradox \cite{huisman2023p,wagenmakers2023history}. Emphasizing effect sizes rather than $p$-value significance thresholds, a quadratic exponential relationship between the Bayes factor and the separation of credible intervals is described in \cite{wei2023investigating}. However, computation remains a practical hurdle: Bayes factors require numerical integration over prior distributions, often via Markov chain Monte Carlo (MCMC) with model-specific tuning \cite{JASP2025, BF_Package,sinharay2005empirical}. While this computation is feasible with existing algorithms, a simple Bayes factor calculation that requires only the $p$-value, sample size, and dimension of the test has greater applicability to situations where the full dataset is unavailable.

    With an emphasis on simplicity and practicality, several approximate Bayes factors based on frequentist calculations have been proposed. For example, certain Bayes factors, with informed priors on a focal parameter, yield fast approximations and are based on Jeffreys’s approximate objective Bayes factor (JAB) \cite{jeffreys1935some,jeffreys1936further,bartovs2023general,rostgaard2023simple}. The Bayesian information criterion (BIC) has been extensively used to approximate the Bayes factor \cite{kass1995reference, bickel2025small}. Minimum-Bayes-factor calibrations provide conservative evidence assessments when only a $p$-value is available \cite{held2018p}. Bayes factors based on parametric test statistics are developed by \cite{johnson2005bayes}, with model-selection consistency under the prior predictive and connections to BIC examined in \cite{johnson2008properties}; extensions to nonparametric statistics appear in \cite{yuan2008bayesian}. A proposed $3p\sqrt{n}$ rule, based on a piecewise approximation to the $\chi^{2}_{1}$ quantile function, facilitates JAB in the scalar-parameter case \cite{wagenmakers2022approximate}. We define $\mathit{eJAB}$, a generalization of JAB that incorporates a finite-sample correction and explicitly accounts for the dimension of the test. It is a one-line calculation that is model-selection consistent under broad conditions. $\mathit{eJAB}$ reduces to JAB when the parameter under test is a scalar and as $n\to\infty$.

    Our work differs from \cite{bickel2025small} in its consideration of $p$-values beyond those from the likelihood ratio test and in our explicit statement of conditions for model-selection consistency. For the latter, \cite{bickel2025small} demonstrates asymptotic equivalence to the BIC so that model-selection consistency arises under the conditions specified in \cite{neath2012bayesian}. It differs from \cite{bartovs2023general,rostgaard2023simple,wagenmakers2022approximate} primarily in our formulation of the generalization and in our development of a theoretical justification.
    
    Our contributions are thus threefold. First, we define an extension of JAB, $\mathit{eJAB}$, which generalizes to multidimensional parameters by incorporating the dimension. Our approach differs from BIC-based approximations by accommodating $p$-values beyond those from likelihood-ratio tests, and from other methods by establishing a model-selection consistency justification under regularity conditions. Second, we assess the finite-sample performance of $\mathit{eJAB}$ across ten common parametric and nonparametric tests. The results suggest that $\mathit{eJAB}$ is a broadly applicable and reliable approach to approximate the Bayes factor. Third, we apply $\mathit{eJAB}$ to 71,126 reported findings from \url{https://ClinicalTrials.gov} (CTG), accessed on August 7th, 2025, and argue that NHST ($p$-value $\leqslant\alpha$) and $\mathit{eJAB}$ ($>1$, favoring the null) contradictions are candidate type I errors. We flag 4,088 such results and find that the distribution of statistical evidence is uniform across five clinical trial phases for both primary and secondary outcomes in CTG. We also identify 487 instances of the Jeffreys–Lindley paradox.

\section{Methodology}

\begin{definition} \label{eJAB}

To test the hypotheses $\mathcal{H}_0:\boldsymbol{\theta}=\boldsymbol{\theta}_0$ versus $\mathcal{H}_1:\boldsymbol{\theta}\neq\boldsymbol{\theta}_0$, we define $\mathit{eJAB}$ as

\begin{equation*}
  \mathit{eJAB}_{01}=\sqrt{n}\exp\left\{-\frac{1}{2}\frac{n^{1/q}-1}{n^{1/q}}\ \!Q_{\chi^{2}_{q}}(1-p)\right\},
\end{equation*}
where $q$ is the dimension of the test, $n$ is the sample size, $Q_{\chi^{2}_{q}}(\cdot)$ is the quantile function of the chi-squared distribution with $q$ degrees of freedom (df), and $p$ is the $p$-value from an NHST. We further define $\mathit{eJAB}_{10} = 1\ \!/\ \!\mathit{eJAB}_{01}$.

\end{definition}

\begin{lemma} \label{ineq}
For $x\to1^-$,
\begin{equation*} 
  0<f(x)=-2\ln(1-x)-2\ln(-\ln(1-x))<Q_{\chi^{2}_{q}}(x).
\end{equation*}
The proof is provided on the Open Science Framework at \url{https://osf.io/qeprj/} (see files within `Theorem').
\end{lemma}

\begin{theorem} \label{thm}
Assuming the NHST $p$-value satisfies two regularity conditions,

\begin{itemize}
    \item \textbf{R1}: $p\overset{D}{\to}\operatorname{Unif}(0,1)$ as $n\to\infty$  under $\mathcal{H}_0$,
    \item \textbf{R2}: $D_{n}=-\sqrt{n}\cdot p\ln{p}\overset{P}{\to}0$ as $n\to\infty$ under $\mathcal{H}_1$,
\end{itemize}
then,
\begin{equation*} 
\plim_{n\to\infty}\mathit{eJAB}_{01}=
\begin{cases}
    \ \infty & \text{under}\ \mathcal{H}_0, \\
    \ 0 & \text{under}\ \mathcal{H}_1.
\end{cases}
\end{equation*}

\noindent \textit{Proof}: 

Assume $\mathcal{H}_0$ is true. Let $\delta\in(0,1)$ and $K>0$ be arbitrary. Note that \textbf{R1} and $W_n=Q_{\chi^{2}_{q}}(1-p) \overset{D}{\to} \chi^{2}_{q}$ by the continuous mapping theorem. For $n\geqslant N^*=\lceil K^2\exp\{Q_{\chi^{2}_{q}}(1-\delta)\}\rceil$, we have

\begin{align*}
  &\mathbb{P}(\mathit{eJAB}_{01}>K) \\
  =&\mathbb{P}\left(\sqrt{n}\exp\left\{-\frac{1}{2}\frac{n^{1/q}-1}{n^{1/q}}\ \!W_n\right\}>K\right) \\
  >& \mathbb{P}\left(\sqrt{n}\exp\left\{-\frac{1}{2}\ \!W_n\right\}>K\right) \\
  =&\mathbb{P}\left(W_n<-2\ln{\frac{K}{\sqrt{n}}}\right) \\
  \geqslant&\mathbb{P}\left(W_n<-2\ln{\frac{K}{\sqrt{N^*}}}\right) \\
  \geqslant&\mathbb{P}\left(W_n<-2\ln{\frac{K}{\sqrt{K^{2}\exp\{Q_{\chi^{2}_{q}}(1-\delta)\}}}}\right) \\
  =&\mathbb{P}\left(W_n< Q_{\chi^{2}_{q}}(1-\delta)\right) \\
  =&1-\delta.
\end{align*}

Thus, $\mathit{eJAB}_{01}\overset{P}{\to}\infty$ as $n\to\infty$ under $\mathcal{H}_0$.

Assume $\mathcal{H}_1$ is true. Let $\delta\in(0,1)$ and $\epsilon>0$ be arbitrary. Lemma~\ref{ineq} gives $0<f(1-p)<Q_{\chi^{2}_{q}}(1-p)$ for $p\to0$.
\begin{align*}
  &\mathbb{P}(\mathit{eJAB}_{01}<\epsilon) \\
  =&\mathbb{P}\left(\sqrt{n}\exp\left\{-\frac{1}{2}\frac{n^{1/q}-1}{n^{1/q}}\ \!Q_{\chi^{2}_{q}}(1-p)\right\}<\epsilon\right) \\
  >&\mathbb{P}\left(\sqrt{n}\exp\left\{-\frac{1}{2}\frac{n^{1/q}-1}{n^{1/q}}(-2\ln{p} -2\ln(-\ln{p)})\right\}<\epsilon\right) \\
  =&\mathbb{P}\left(\sqrt{n}\ \left(p\cdot(-\ln{p})\right)^{\frac{n^{1/q}-1}{n^{1/q}}}<\epsilon\right) \\
  >&1-\delta.
\end{align*}

The final inequality is true for $n$ sufficiently large by \textbf{R2} and the continuous mapping theorem with $1-\frac{1}{n^{1/q}}\to1$.
 
Thus, $\mathit{eJAB}_{01}\overset{P}{\to}0$ as $n\to\infty$ under $\mathcal{H}_1$. Together, $\mathit{eJAB}$ is model-selection consistent.

$\hfill\blacksquare$
\end{theorem}

\begin{remark}
We can derive $\mathit{eJAB}$ as an approximation using a generalization of the multivariate normal unit-information prior for the parameters under test,
\begin{equation*}
    \boldsymbol{\theta}\sim\boldsymbol{\mathcal{N}}_q\!\left(\hat{\boldsymbol{\theta}},\ n^{1/q}\cdot \mathbf{J}^{-1}_n(\hat{\boldsymbol{\theta}})\right),
\end{equation*}
where $\hat{\boldsymbol{\theta}}$ is the maximum likelihood estimate (MLE), and $\mathbf{J}_n(\hat{\boldsymbol{\theta}})$ is the observed information matrix of size $q\times q$. Under an asymptotic Gaussian approximation, the posterior density takes the form,
\begin{equation*}
    \pi(\boldsymbol{\theta}\mid\boldsymbol{y},\mathcal{H}_1)\approx(2\pi)^{-\frac{q}{2}}\cdot\lvert \mathbf{J}_n(\hat{\boldsymbol{\theta}})\rvert^{\frac{1}{2}}\cdot\exp{\left\{-\frac{1}{2}(\boldsymbol{\theta}-\hat{\boldsymbol{\theta}})^\top\cdot \mathbf{J}_n(\hat{\boldsymbol{\theta}})\cdot(\boldsymbol{\theta}-\hat{\boldsymbol{\theta}})\right\}}.
\end{equation*}

Assuming that the prior density for any nuisance parameters under $\mathcal{H}_0$ is everywhere equal to the conditional prior density for those parameters under $\mathcal{H}_1$, given the null value of the parameter under test, the Bayes factor (BF) can be expressed as the Savage–Dickey density ratio \cite{mulder2022generalization},
\begin{equation*}
    \mathit{BF}_{01}=\frac{\pi(\boldsymbol{\theta}_0\mid\boldsymbol{y},\mathcal{H}_1)}{\pi(\boldsymbol{\theta}_0\mid \mathcal{H}_1)}\approx\sqrt{n}\exp\left\{-\frac{1}{2}\frac{n^{1/q}-1}{n^{1/q}} W\right\},
\end{equation*}
where $\boldsymbol{y}$ represents the data, and $W$ is the Wald statistic. When the $p$-value is obtained from a Wald test, substituting $W=Q_{\chi^{2}_{q}}(1-p)$ yields $\mathit{eJAB}_{01}$ in Definition~\ref{eJAB}.

The prior underlying the approximate BF interpretation is centered around the MLE $\hat{\boldsymbol{\theta}}$ with $\lvert\operatorname{Var}(\boldsymbol{\theta})\rvert=n\ \!\lvert\mathbf{J}^{-1}_n(\hat{\boldsymbol{\theta}})\rvert$. When $q=1$, this is the unit-information prior. More generally, the volume of prior probability ellipsoids around the MLE is $\mathcal{O}(n^{(1-q)/2})$. The prior distribution is asymptotically diffuse relative to the likelihood.
\end{remark}

\begin{remark}
When the $p$-value is obtained from a likelihood-ratio test and $q=1$, $\mathit{eJAB}$ reduces to the evidential BIC method of \cite{bickel2025small}. This approximation is derived by applying a finite-sample correction to the BIC approximation to the BF, which amounts to raising the likelihood ratio to the power $1-\frac{1}{n}$. Model-selection consistency for the evidential BIC follows from its asymptotic equivalence to BIC together with standard consistency results for BIC \cite{bickel2025small,neath2012bayesian}. By contrast, the conditions of Theorem~\ref{thm} are broadly applicable: they cover nonparametric tests, provided \textbf{R1} and \textbf{R2} hold.
\end{remark}

\section{Simulation Studies}

We conduct 12 simulation studies across ten parametric and nonparametric tests to evaluate the regularity conditions and finite-sample accuracy. For each design, 5,000 datasets are generated under the null hypothesis and the alternative hypothesis with small, medium, and large effect sizes across various sample sizes. The designs include:
\begin{enumerate}
\item Two-sample $t$-test;
\item Simple linear regression;
\item Simple logistic regression;
\item One-way analysis of variance (ANOVA);
\item One-way repeated-measures ANOVA (rANOVA) with high and low intraclass correlations $\rho\in\{.9,\ \!.2\}$;
\item Chi-squared tests for independence under multinomial (total count fixed) and product-multinomial (row counts fixed) designs;
\item Cox proportional-hazards regression with right censoring;
\item Wilcoxon signed-rank test;
\item Mann–Whitney $U$-test;
\item Kruskal–Wallis $H$-test.
\end{enumerate}
For nonparametric tests (items 8 through 10), we simulated using right-skewed and heavy-tailed (non-Gaussian) distributions. Detailed simulation settings, additional designs (e.g., two-way ANOVA and large contingency tables), and the \texttt{R} code for the subsequent subsections are available on the Open Science Framework at \url{https://osf.io/qeprj/} (see files within `Theorem', `Simulations' [parametric], and `Supplementary Information' [nonparametric]).

We used $p$-values from the $t$-test for items 1 and 2; the $F$-test for items 4 and 5; the Wald test for items 3 and 7; and the asymptotic null distribution for item 6. The $p$-values for 8–10 are based on asymptotic distributions. The effective sample size $n$ was taken as the number of uncensored observations for survival analysis and as the number of independent observations (i.e., the product of the number of participants and one less than the number of conditions) for rANOVA \cite{nathoo2016bayesian,masson2011tutorial}. For the other designs, $n$ always refers to the total number of observations or counts. In particular, we tested the harmonic mean of group sizes for item 1, as in JAB \cite{wagenmakers2022approximate}, but do not demonstrate it here. Among the tests with multiple \textit{df} ($q>1$) are items 4 through 6, and 10.

\begin{figure*}[!htb]
\centering
\includegraphics[width=0.9\textwidth]{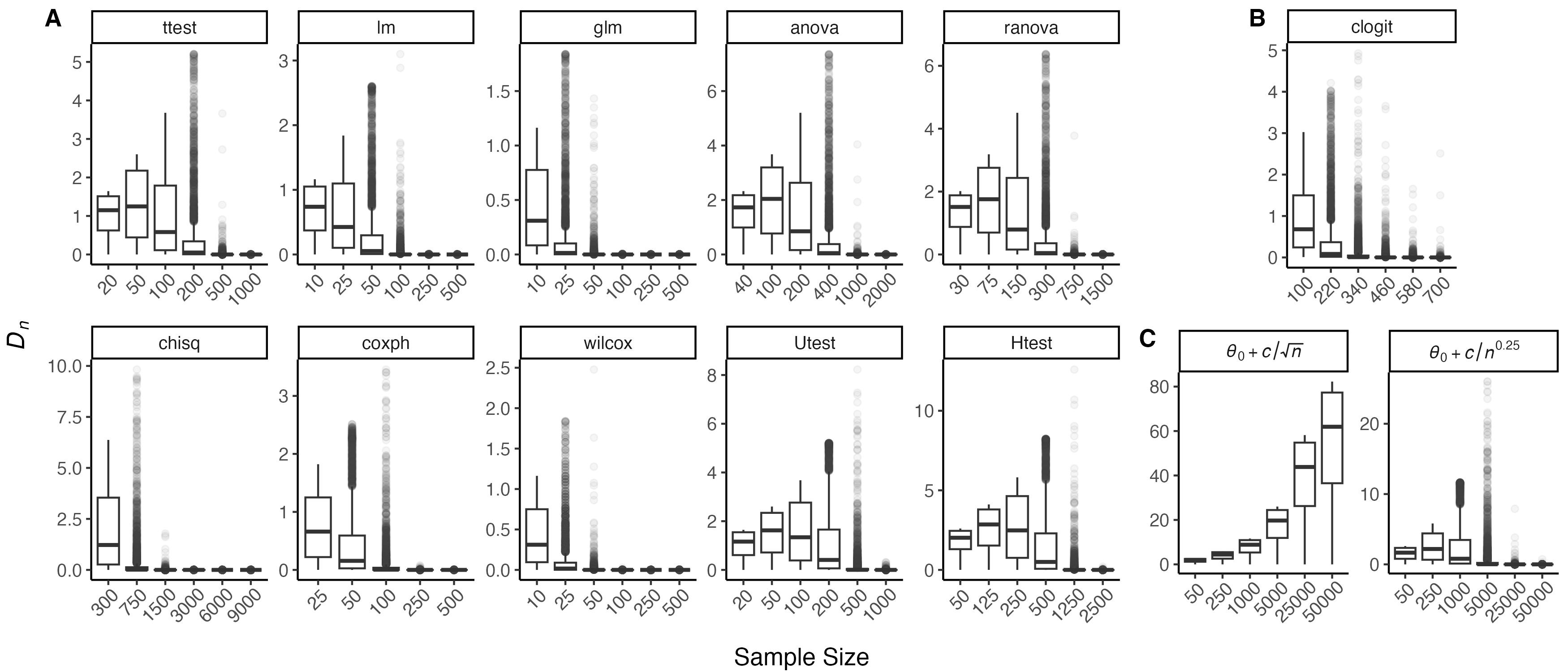}
\caption{Sampling distributions of $D_{n}=-\sqrt{n}\cdot p\ln{p}$ under (A–B) the fixed and (C) the local $\mathcal{H}_1$ for eleven statistical tests. Panel labels: ttest ($t$-test), lm (linear regression), glm (logistic regression), anova (analysis of variance), ranova (repeated-measures ANOVA), chisq (chi-squared test), coxph (Cox proportional-hazards regression), wilcox (Wilcoxon signed-rank test), Utest (Mann–Whitney test), Htest (Kruskal–Wallis test), and clogit (conditional logistic regression).}
\label{fig1}
\end{figure*}

\begin{figure*}[!htb]
\centering
\includegraphics[width=\textwidth]{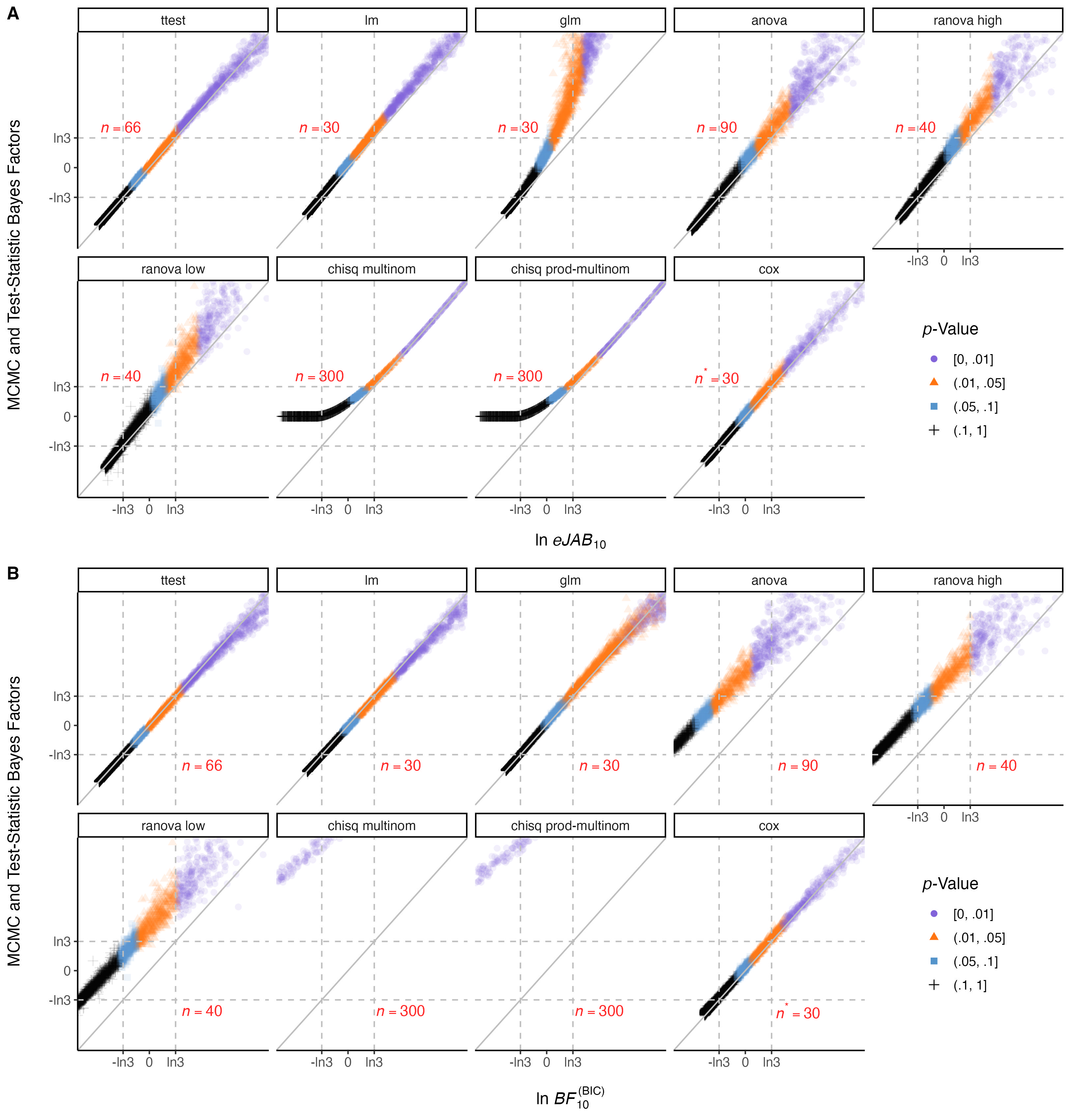}
\caption{Comparison of the natural logarithmic (A) $\mathit{eJAB}_{10}$ and (B) BIC approximation to the Bayes factor with alternative Bayes-factor computations at a fixed sample size. The $y$-axis shows the MCMC-based Bayes factor for all panels, except for the two chi-squared tests, which use the test-statistic Bayes factor instead. The red text indicates the effective sample size $n$, except for the Cox model, which shows the number of observations. The dashed lines indicate $\mathit{eJAB}_{10}$ of 1/3 and 3, representing moderate evidence against and for an effect, respectively.}
\label{fig2}
\end{figure*}

\subsection{Verifying the Regularity Conditions}

Condition \textbf{R1} of Theorem~\ref{thm} is satisfied whenever the $p$-value is computed from an exact or asymptotic null distribution with a continuous cumulative distribution function.

Next, we computed $D_n$ of \textbf{R2} using the $p$-values and sample sizes from the simulated data, generated with medium effect sizes, across the ten designs, plus conditional logistic regression used in our application. In this subsection, items 4 and 5 ($\rho=.9$) each have $4-1=3$ \textit{df}; item 6 has $(3-1)\times(3-1)=4$ \textit{df} under a multinomial design; item 7 represents a semi-parametric Cox model; item 10 has $5-1=4$ \textit{df}. Due to randomness, the $x$-axis has to display the total number of observations or pairs, rather than the number of uncensored observations for item 7, or the number of matched pairs for conditional logistic regression, as required in $D_{n}$ and $\mathit{eJAB}$. Condition \textbf{R2} appears to hold for all fixed alternatives we examined (Fig.~\ref{fig1}A–B). For local alternatives of the form $\theta^{*}=\theta_{0}+c/n^{\kappa}$ in the one-sample $t$-test, \textbf{R2} does not hold at $\kappa=1/2$ but does hold at $\kappa=1/4$ (Fig.~\ref{fig1}C). This follows from the denominator of the $t$-statistic being $\mathcal{O}_{p}(1/\sqrt{n})$.

\subsection{Accuracy of the Bayes-Factor Approximation}
To assess the finite-sample accuracy of an approximation to the BF, we compared $\mathit{eJAB}$ with alternative computations in nine studies based on the parametric designs listed in items 1–7, using fixed representative sample sizes. In this subsection, items 4 and 5 each have $3-1=2$ \textit{df}; item 6 maintains 4 \textit{df}; item 7 now represents a parametric survival regression model with an exponential baseline and expected censoring fractions of 23\% under the null, and 21\%, 19\%, and 18\% under increasing effect sizes.

Except for the two chi-squared tests, the MCMC-based BFs were computed by sampling the marginal posterior density using `\textit{rstan}' \cite{stan}, applying logspline density estimation \cite{kooperberg1992logspline}, and calculating the Savage–Dickey density ratio. In these cases, the model underlying the MCMC sampler assumed the same prior as $\mathit{eJAB}$, so discrepancies reflect errors of approximation rather than differences in prior. For the chi-squared tests, we compared $\mathit{eJAB}$ with (i) test-statistic BFs \cite{johnson2005bayes} and (ii) Dirichlet BFs from \texttt{contingencyTableBF} in the `\textit{BayesFactor}' \texttt{R} package \cite{BF_Package,gunel1974bayes}.

Pairs of $(\ln\mathit{eJAB}_{10},\ 
\ln\mathit{BF}_{10})$ are color- and shape-coded by the $p$-value. Overall, $\mathit{eJAB}$ tracks the MCMC-based BFs closely (Fig.~\ref{fig2}A). Some inaccuracy, $\ln\mathit{eJAB}_{10}< 
\ln\mathit{BF}_{10}$, appears for (binary) logistic regression when evidence favors $\mathcal{H}_1$. For chi-squared tests, agreement with test-statistic BFs is excellent when evidence favors $\mathcal{H}_1$. As evidence shifts toward $\mathcal{H}_0$, $\mathit{eJAB}_{10}$ decreases toward zero, as expected under model-selection consistency, whereas the test-statistic BF is inherently bounded below by 1. Additional comparisons with Dirichlet BFs for chi-squared tests are provided in \textit{Appendix B}. The two Bayes factors reach broad agreement, although differences persist at larger $n$ (\textit{Appendix B}, Fig.~\ref{fig6}). At $n=1,500$, under $\mathcal{H}_0$, the Dirichlet $\mathit{BF}_{10}<1/3$ is almost always true, whereas $\mathit{eJAB}_{10}<3$ is mostly observed; under $\mathcal{H}_1$, $\mathit{eJAB}_{10}$ more readily identifies an effect. Replacing $\mathit{eJAB}$ with the BIC approximation to the BF performs well when $q=1$, but marked severe discrepancies for $q>1$, indicating that $\mathit{eJAB}$ better approximates BF for ANOVA, rANOVA, and chi-squared tests (Fig.~\ref{fig2}B).

In sum, $\mathit{eJAB}$ provides an accurate approximation to Bayes factors across a range of designs and effect sizes for parametric tests.

\subsection{Bayesian Hypothesis Tests Using Nonparametric Statistics}

Classical nonparametric tests do not specify a sampling distribution for the data, so marginal likelihoods are not defined. As such, the Bayes factor can be defined based on the marginal likelihood of the test statistic rather than on the data \cite{yuan2008bayesian}. We compared $\mathit{eJAB}$ with Bayes factors constructed from nonparametric statistics ($\mathit{BFNP}$) across items 8 through 10. The supplementary material reports sequences of box plots depicting how the sampling distributions of $\mathit{eJAB}$ and $\mathit{BFNP}$ vary with effect size and sample size. Figures depicting these results, a detailed description of the experiments, and the associated \texttt{R} Markdown files are available on the Open Science Framework at \url{https://osf.io/qeprj/} (see the `Supplementary Information' directory).

Under $\mathcal{H}_1$, when effects and samples are both relatively small, $\mathit{BFNP}$ tends to provide stronger evidence for $\mathcal{H}_1$ than $\mathit{eJAB}$. With larger effects or larger samples, the two methods produce comparable results. Under $\mathcal{H}_0$, $\mathit{eJAB}$ effectively quantifies evidence for $\mathcal{H}_0$, as with chi-squared tests.

\section{Detecting Type I Errors in 71,126 Clinical Trial Findings with eJAB}

\begin{figure*}[!htb]
\centering
\includegraphics[width=\textwidth]{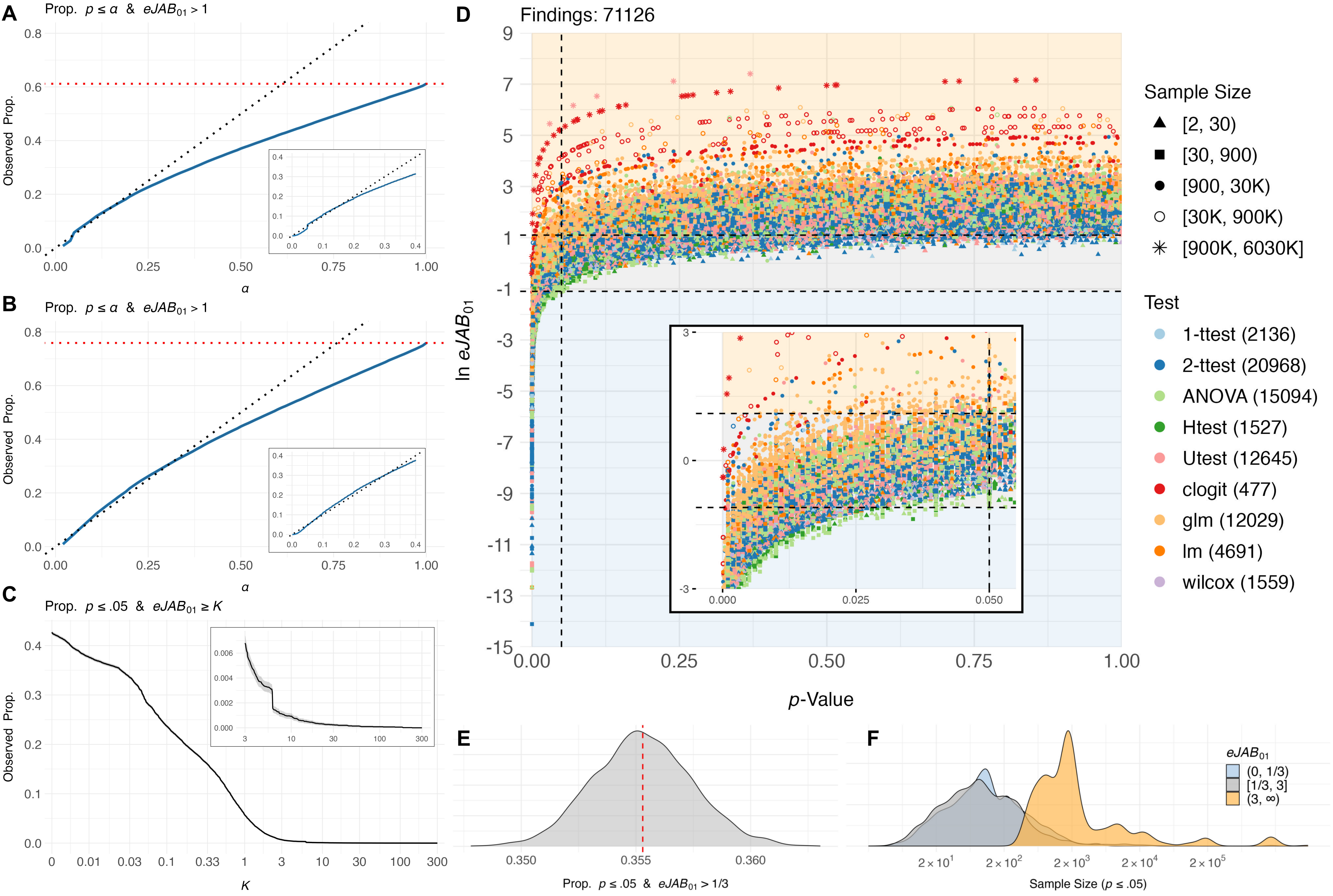}
\caption{Evidence in 71,126 results. (A–B) Proportion of analyses with $p\leqslant\alpha$ and $\mathit{eJAB}_{01}>1$ as candidates type I errors, with $p=.05$ excluded in (B). The red dashed line marks the overall share with $\mathit{eJAB}_{01} > 1$ as $\alpha \rightarrow 1$. (C) Proportion of $p\leqslant.05$ with Bayes-factor threshold $K$. The inset highlights the Jeffreys–Lindley paradox region. (D) Scatterplot of $\ln \mathit{eJAB}_{01}$ versus $p$-value, with an inset zoom for $p\leqslant.05$. Shaded bands denote evidence categories. (E) Posterior distribution of the proportion of $p\leqslant.05$ and $\mathit{eJAB}_{01}>1/3$. The red dashed line indicates the posterior mean. (F) Density of sample sizes for all $p\leqslant.05$, stratified by Bayes-factor category.}
\label{fig3}
\end{figure*}

Having evaluated $\mathit{eJAB}$ across a wide range of tests, we apply it to reevaluate the evidence in a large corpus of CTG results. Prior work has raised concerns about the strength of statistical evidence in clinical research. For example, a 2012 study reported replication for only 11\% of landmark preclinical cancer findings \cite{Begley2012-ur}. A 2016 study found that most late-stage trials fail, largely for lack of efficacy \cite{Hwang2016-pw}. And, a 2019 study estimated that 90\% of drug-development trials fail for similar reasons \cite{Sun2022-ma}. We posit that failures at later phases may reflect type I errors, where the null hypothesis was incorrectly rejected at an earlier phase.

Our objective is to characterize the overall strength of evidence in 71,126 CTG findings and to identify potential type I errors. These results span 43 MeSH categories (e.g., neoplasms, cardiovascular diseases, and autoimmune diseases) and encompass nine statistical tests: one-sample $t$-test, two-sample $t$-test, linear regression, logistic regression, ANOVA, conditional logistic regression, Wilcoxon signed-rank test, Mann–Whitney test, and Kruskal–Wallis test. We verify that regularity condition \textbf{R2} holds for conditional logistic regression (Fig.~\ref{fig1}C).

For each, we extracted the $p$-value, sample size, and parameter dimension, and, thus, computed $\mathit{eJAB}_{01}$. We excluded survival analyses and chi-squared tests because CTG usually omits the number of uncensored observations and full contingency-table dimensions. Preprocessing to address misreported and left-censored $p$-values, sample sizes, and test labels \cite{Miron2020-gm} is described in \textit{Appendix A}.

The relationship between $\alpha$ and the share of reported findings with $p\leqslant\alpha$ and $\mathit{eJAB}_{01}>1$ is strongly linear for $\alpha\in(0,.25)$, with a visible jump at $\alpha=.05$ due to results reported exactly as $p=.05$ (Fig.~\ref{fig3}A). We attribute this to left-censoring at the target threshold. Restricting to complete cases removes this artifact and yields an almost perfectly linear trend for $\alpha\in(0,.4)$, strongly suggesting that findings with $p\leqslant\alpha$ and $\mathit{eJAB}_{01}>1$ are type I errors at level $\alpha$ (Fig.~\ref{fig3}B). 

We therefore classify findings with $p\leqslant\alpha$ and $\mathit{eJAB}_{01}>1$ as candidate type I errors at the significance level $\alpha$. Among 30,790 results significant at $\alpha=.05$, we catalog 4,088 candidates. Although lower thresholds have been recommended in the literature \cite{benjamin2018redefine, johnson2013revised}, $\alpha=.05$ remains germane in CTG: in a random sample of 52 trial protocols, 75\% (SE 6\%) explicitly set $\alpha\geqslant.05$ as the target evidential threshold. We list all 4,088 candidates on the Open Science Framework at \url{https://osf.io/qeprj/} (see `Supplementary Information') and provide an annotated subset of 131 cases with $p\leqslant.01$ and $\mathit{eJAB}_{01}>1$, cross-referenced with CTG and PubMed. 

We fit a hierarchical Dirichlet-multinomial model with a flat prior on study-specific proportions of $p\leqslant.05$ and $\mathit{eJAB}_{01}>1/3$, aggregating draws to estimate the overall proportion. We estimate that 35.5\% (posterior SD 0.22\%) of $p\leqslant.05$ results correspond to $\mathit{eJAB}_{01}>1/3$, i.e., evidence for the alternative no stronger than anecdotal (Fig.~\ref{fig3}E).

In the plot of $\ln\mathit{eJAB}_{01}$ versus $p$, with points colored by test type and shaped by sample size, the region to the left of $p=.05$ and above $\mathit{eJAB}_{01}=3$ indicates instances of the Jeffreys–Lindley paradox—that is, cases where the $p$-value leads to rejection of $\mathcal{H}_0$, while the Bayes factor favors $\mathcal{H}_0$ over $\mathcal{H}_1$. Notably, $\ln \mathit{eJAB}_{01}$ shows substantial variability at the smallest $p$-values (Fig.~\ref{fig3}D). Contradictions with $p \leqslant .05$ yet $\mathit{eJAB}_{01}>3$ occur predominantly at larger sample sizes, where observed effect sizes can be negligible (Fig.~\ref{fig3}F).

We also examine the proportion of all $p\leqslant.05$ and $\mathit{eJAB}_{01}\geqslant K$ for $K\in(0,300)$; at $K=1$, this proportion recovers the candidate type I error at $\alpha=.05$. The inset highlights Jeffreys–Lindley paradox cases (Fig.~\ref{fig3}C). We identify 487 such instances in the CTG data. Additional results stratified by MeSH category appear in \textit{Appendix B} (Figs.~\ref{fig8}–\ref{fig10}).

In sum, the near-perfect linearity indicates that findings with $p\leqslant\alpha$ and $\mathit{eJAB}_{01}>1$ behave as type I errors in the CTG data when tested at $\alpha\in(0,.2]$. To elucidate how such candidates arise, consider a Wald test with $q=1$ and independent and identically distributed data for fixed $n<\infty$ and $\alpha>1-F_{\chi^{2}_{1}}\left(\frac{n \ln{n}}{n-1}\right)$. A candidate type I error occurs when the observed effect size falls within the interval,
\begin{equation*}
\sqrt{\frac{1}{n}\ \!Q_{\chi^{2}_{1}}(1-\alpha)}\cdot I_{1}^{-\frac{1}{2}}(\hat{\theta}) \leqslant \lvert\hat{\theta} - \theta_{0}\rvert < \sqrt{\frac{\ln n}{n-1}}\cdot I_{1}^{-\frac{1}{2}}(\hat{\theta}),
\end{equation*}
where $I_{1}$ is the Fisher information for one observation. In this regime, the observed effect size is large enough to achieve standard statistical significance at sample size $n$. Small enough, $eJAB_{01}$ favors the null. To favor the alternative, the latter requires the observed effect size to exceed a bound of order $\mathcal{O}_{p}\left(\sqrt{\frac{\ln n}{n}}\right)$. 

We review several primary outcomes from interventional clinical trials designated as candidate type I errors at $\alpha=.01$.

\begin{enumerate}[leftmargin=1.5cm, rightmargin=1.5cm,itemsep=1em]
    \item \textbf{\#NCT02605174.~Three doses of lasmiditan (50, 100, and 200 mg) compared with placebo for the acute treatment of migraine:} $p=.009$ and $\mathit{eJAB}_{01}=1.06$. Phase 3 trial of lasmiditan for acute migraine. We flag as a candidate type I error the reported effect that 50 mg increased the odds of being free of the most bothersome symptom versus placebo. 
    
    \item \textbf{\#NCT00337727.~Aprepitant for the prevention of chemotherapy-induced nausea and vomiting:} $p=.01$ and $\mathit{eJAB}_{01}=1.05$. Phase 3 trial in patients starting moderately emetogenic chemotherapy. We flag as a candidate type I error the reported effect that aprepitant increased the odds of no vomiting versus standard therapy.
    
    \item \textbf{\#NCT01087541.~Evaluation of safety in patients with diabetes:} $p=.01$ and $\mathit{eJAB}_{01}=3.00$. Trial of a primary-care professional training program on diabetes endpoints. We flag as a candidate type I error the reported effect that the intervention lowered glycated hemoglobin (long-term blood glucose) relative to usual care.
    
    \item \textbf{\#NCT02896400.~Optimizing tobacco-dependence treatment in the emergency department:} $p=.01$ and $\mathit{eJAB}_{01}=1.18$. Trial of common interventions for adult smokers presenting to the emergency department. We flag as a candidate type I error the reported effect that a brief negotiated interview increased abstinence versus no such interview.     

    \item \textbf{\#NCT02266277.~System alignment for vaccine delivery:} $p=.0026$ and $\mathit{eJAB}_{01}=1.07$. Trial of nonmedical strategies to increase influenza and pneumococcal vaccination. We flag as a candidate type I error the reported effect that a patient-portal vaccine reminder increased completion versus no message.
\end{enumerate}

\begin{figure*}[!htb]
\centering
\includegraphics[width=0.8\textwidth]{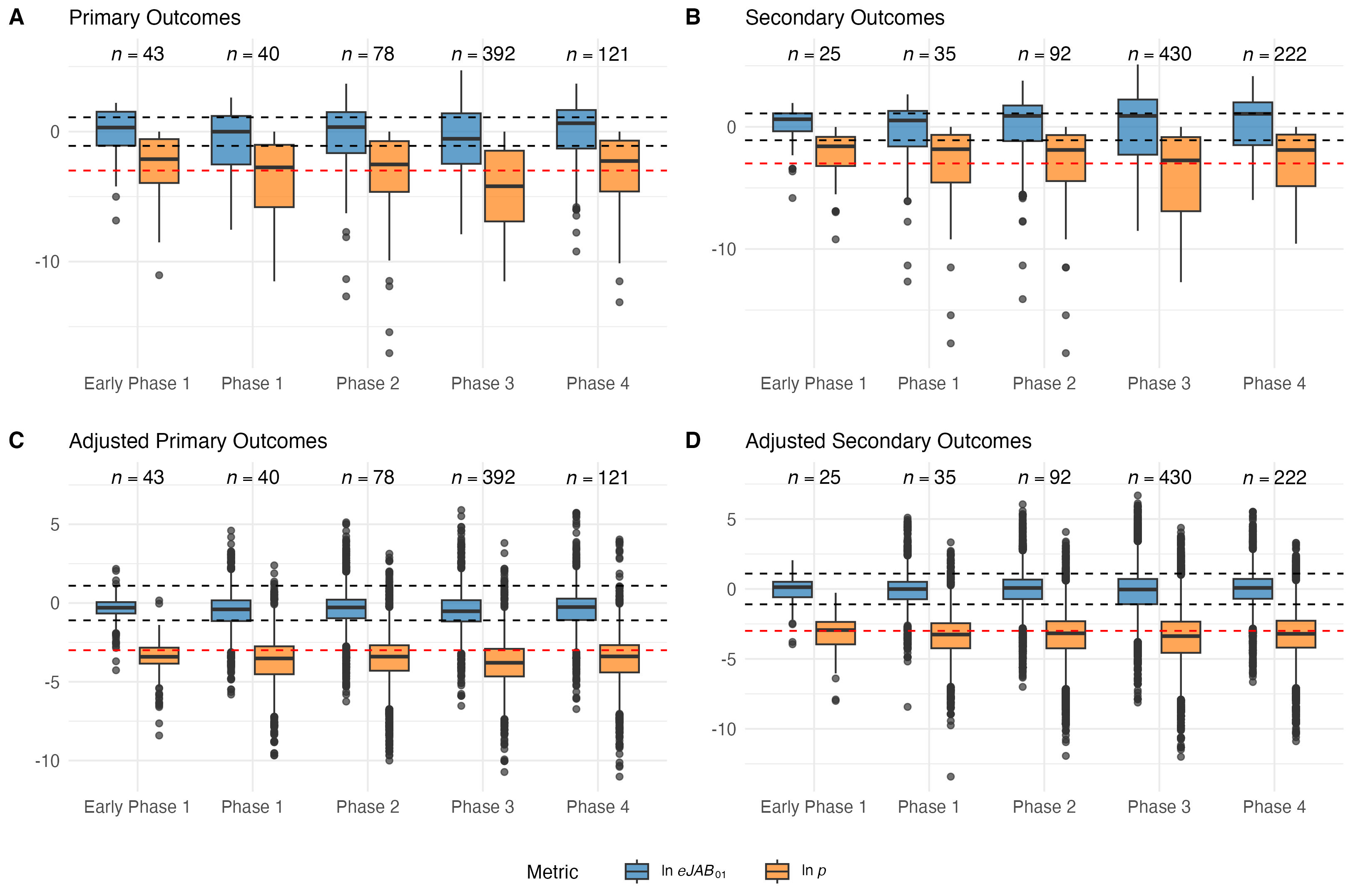}
\caption{Measures of evidence by clinical trial phase. (A–B) Distributions of $\ln\mathit{eJAB}_{01}$ and $\ln{p}$ by phase for primary (A) and secondary (B) outcomes. (C–D) Distributions of adjusted (A–B). Adjusted measures are obtained from linear mixed models. The average analysis sample size is shown above each facet. Black dashed lines mark $-\ln3$ and $\ln3$. The red dashed line marks $\ln0.05$.}
\label{fig4}
\end{figure*}

\begin{figure*}[!htb]
\centering
\includegraphics[width=4.2in,height=3.4in]{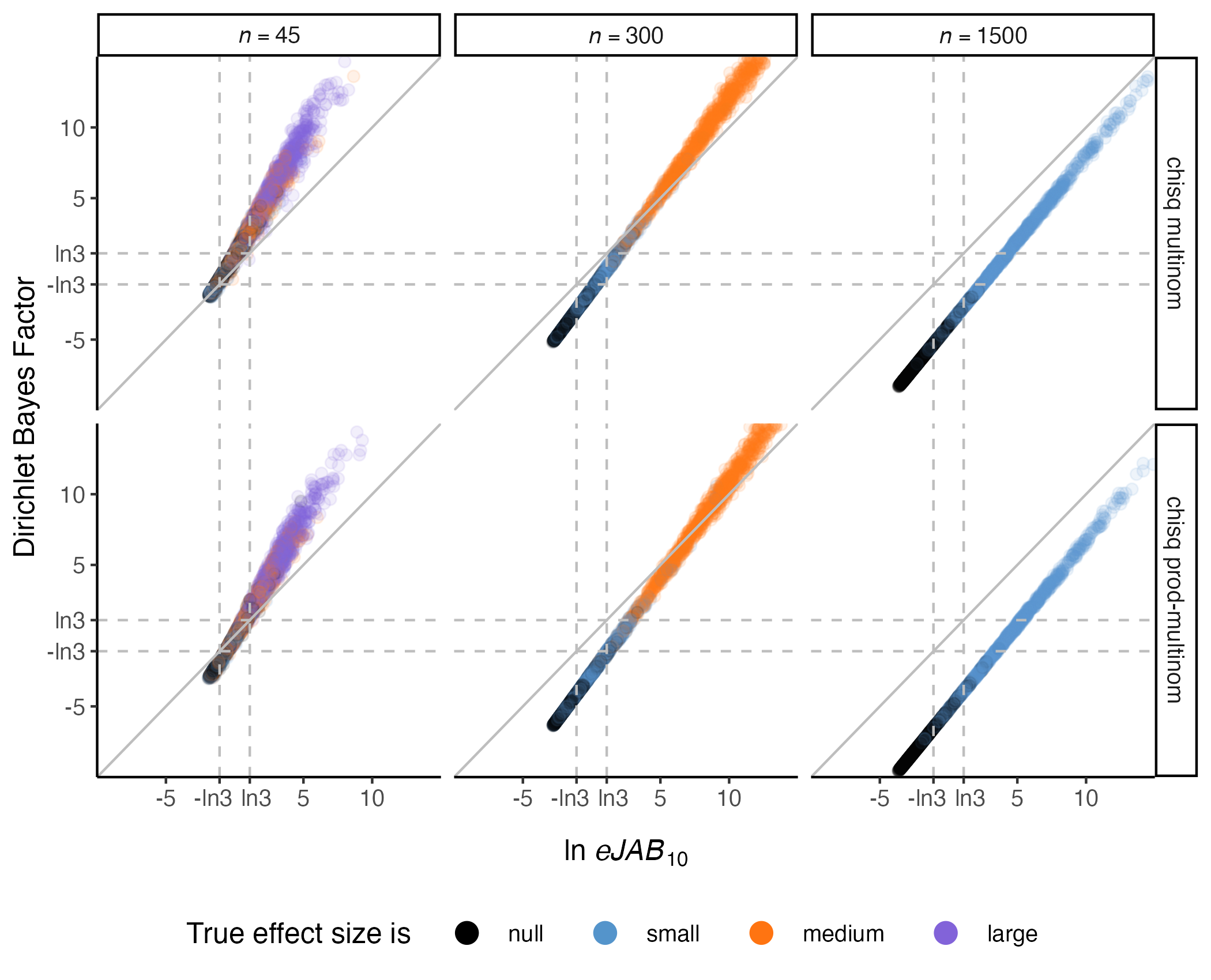}
\caption{Comparison of the natural logarithmic $\mathit{eJAB}_{10}$ with the \texttt{contingencyTableBF} function from the `\textit{BayesFactor}' \texttt{R} package at three total counts for chi-squared tests of independence under two sampling designs: multinomial (top) and product-multinomial (bottom). Colors represent the true effect size. The dashed lines indicate $\mathit{eJAB}_{10}$ of 1/3 and 3, representing moderate evidence against and for an effect, respectively.}
\label{fig6}
\end{figure*}

We examine the distributions of $\ln\mathit{eJAB}_{01}$ and $\ln{p}$ by trial phase for primary and secondary outcomes (Fig.~\ref{fig4}A–B). For primary outcomes, there is no discernible trend toward stronger evidence from early phase to phase 4. The distributions of $\ln{p}$ shift downward at phase 3, consistent with the markedly larger average sample sizes at that phase. For secondary outcomes, we again observe no phase-wise increase in evidence, though variability in both measures is higher in phase 3. Plotting $p$ and sample size by phase shows that changes in $p$ track changes in sample size (\textit{Appendix B}, Fig.~\ref{fig12}).

To better visualize trends in evidence across phases, we fit a linear mixed model with a fixed intercept and trial-specific random effects to each of $\ln\mathit{eJAB}_{01}$ and $\ln{p}$, and then examine residuals re-centered by the estimated intercept (\textit{Appendix B}, Fig.~\ref{fig11}). The adjusted measures are essentially uniform across phases (Fig.~\ref{fig4}C–D). For primary outcomes, most adjusted $p$-values lie near~.05 and above~.01. Using $\mathit{eJAB}$, this interpretation corresponds to `anecdotal evidence' or `evidence that is barely worth mentioning' for both primary and secondary outcomes \cite[p.~105]{lee2014bayesian}. Comparing primary with secondary outcomes, we note a slight and phase-invariant shift toward greater evidence for the alternative in primary outcomes.

\section{Conclusions}

$\mathit{eJAB}$ can be interpreted as an approximate objective Bayes factor when $p$-values are obtained from Wald or likelihood-ratio tests. The same interpretation should hold for tests whose statistics differ from the Wald statistic by $o_{p}(1)$. Simulations suggest that this interpretation holds more broadly. In addition, $\mathit{eJAB}$ remains model-selection consistent for $p$-values from any test satisfying \textbf{R1} and \textbf{R2} of Theorem~\ref{thm}. \endnote{All \texttt{R} code and the data used in the analyses are available on the Open Science Framework at \url{https://osf.io/qeprj/}. An $\mathit{eJAB}$ calculator is available at \url{https://flxp0x-puneet-velidi.shinyapps.io/bayesfactor/}.} 

We compared the finite-sample performance of $\mathit{eJAB}$ with Bayes factors computed from MCMC and the Savage–Dickey density ratio for the $t$-test, linear regression, logistic regression, ANOVA, rANOVA, and parametric survival analysis. Under the same generalization of the unit-information prior, $\mathit{eJAB}$ closely approximates the MCMC-based Bayes factor. For a chi-squared test under multinomial or product-multinomial sampling, it agrees with the test-statistic Bayes factor under the alternative, unlike the latter, also registers evidence for the null when true. For nonparametric tests, namely Wilcoxon signed-rank, Mann–Whitney, and Kruskal–Wallis, $\mathit{eJAB}$ performs comparably to Bayes factors based on nonparametric test statistics. Our simulations also show improved performance compared to the BIC approximation to the Bayes factor when $q>1$.

In our reevaluation of 71,126 clinical trials, findings with $p\leqslant\alpha$ and $\mathit{eJAB}_{01}>1$ exhibit characteristics of type I errors in the CTG data when tested at significance level $\alpha\in(0,.2]$. The uniform strength of evidence across phases is consistent with a roughly constant proportion of hypotheses with true effects. The 4,088 CTG candidate type I errors warrant targeted follow-up to investigate replicability.

\bigskip
\bigskip

\section*{Acknowledgements}

This work was supported by a discovery grant to F.S.N.~(RGPIN-04044-2020) from the Natural Sciences and Engineering Research Council of Canada and research accelerator awards to P.V.~and Y.L. from the Maud Menten Institute. P.V.~and Z.W.~contributed equally to this work. Correspondence concerning this paper should be addressed to F.S.N.~(nathoo@uvic.ca).

\vfill\theendnotes

\clearpage

\bibliographystyle{unsrt}  
\bibliography{eJAB}  

\begin{thebibliography}{10}

\bibitem{Ioannidis2005-uq}
John P~A Ioannidis.
\newblock Why most published research findings are false.
\newblock {\em PLOS Medicine}, 2(8):696--701, 2005.

\bibitem{benjamin2018redefine}
Daniel~J Benjamin, James~O Berger, Magnus Johannesson, Brian~A Nosek, Eric-Jan Wagenmakers, Richard Berk, Kenneth~A Bollen, Bj{\"o}rn Brembs, Lawrence Brown, Colin Camerer, et~al.
\newblock Redefine statistical significance.
\newblock {\em Nature Human Behaviour}, 2:6--10, 2018.

\bibitem{johnson2013revised}
Valen~E Johnson.
\newblock Revised standards for statistical evidence.
\newblock {\em Proceedings of the National Academy of Sciences}, 110(48):19313--19317, 2013.

\bibitem{wasserstein2016asa}
Ronald~L Wasserstein and Nicole~A Lazar.
\newblock The {ASA} statement on $p$-values: Context, process, and purpose.
\newblock {\em The American Statistician}, 70(2):129--133, 2016.

\bibitem{berger1987testing}
James~O Berger and Thomas Sellke.
\newblock Testing a point null hypothesis: The irreconcilability of $p$ values and evidence.
\newblock {\em Journal of the American Statistical Association}, 82(397):112--122, 1987.

\bibitem{grunwald2024beyond}
Peter~D Gr{\"u}nwald.
\newblock Beyond {N}eyman--{P}earson: {E}-values enable hypothesis testing with a data-driven alpha.
\newblock {\em Proceedings of the National Academy of Sciences}, 121(39):e2302098121, 2024.

\bibitem{evans2015measuring}
Michael Evans.
\newblock Measuring statistical evidence using relative belief.
\newblock {\em Computational and Structural Biotechnology Journal}, 14:91--96, 2016.

\bibitem{halsey2019reign}
Lewis~G Halsey.
\newblock The reign of the $p$-value is over: What alternative analyses could we employ to fill the power vacuum?
\newblock {\em Biology Letters}, 15(5):1--8, 2019.

\bibitem{kass1995bayes}
Robert~E Kass and Adrian~E Raftery.
\newblock Bayes factors.
\newblock {\em Journal of the American Statistical Association}, 90(430):773--795, 1995.

\bibitem{ly2016evaluation}
Alexander Ly, Josine Verhagen, and Eric-Jan Wagenmakers.
\newblock An evaluation of alternative methods for testing hypotheses, from the perspective of {H}arold {J}effreys.
\newblock {\em Journal of Mathematical Psychology}, 72:43--55, 2016.

\bibitem{raftery1995bayesian}
Adrian~E Raftery.
\newblock Bayesian model selection in social research.
\newblock {\em Sociological Methodology}, 25:111--163, 1995.

\bibitem{morey2016philosophy}
Richard~D Morey, Jan-Willem Romeijn, and Jeffrey~N Rouder.
\newblock The philosophy of {B}ayes factors and the quantification of statistical evidence.
\newblock {\em Journal of Mathematical Psychology}, 72:6--18, 2016.

\bibitem{wagenmakers2007practical}
Eric-Jan Wagenmakers.
\newblock A practical solution to the pervasive problems of $p$ values.
\newblock {\em Psychonomic Bulletin \& Review}, 14(5):779--804, 2007.

\bibitem{kass1995reference}
Robert~E Kass and Larry Wasserman.
\newblock A reference {B}ayesian test for nested hypotheses and its relationship to the {S}chwarz criterion.
\newblock {\em Journal of the American Statistical Association}, 90(431):928--934, 1995.

\bibitem{huisman2023p}
Leendert Huisman.
\newblock Are $p$-values and {B}ayes factors valid measures of evidential strength?
\newblock {\em Psychonomic Bulletin \& Review}, 30(3):932--941, 2023.

\bibitem{wagenmakers2023history}
Eric-Jan Wagenmakers and Alexander Ly.
\newblock History and nature of the {J}effreys--{L}indley paradox.
\newblock {\em Archive for History of Exact Sciences}, 77:25--72, 2023.

\bibitem{wei2023investigating}
Zhengxiao Wei, Farouk~S Nathoo, and Michael E~J Masson.
\newblock Investigating the relationship between the {B}ayes factor and the separation of credible intervals.
\newblock {\em Psychonomic Bulletin \& Review}, 30(5):1759--1781, 2023.

\bibitem{JASP2025}
{JASP Team}.
\newblock {JASP (Version 0.95.2)[Computer software]}, 2025.

\bibitem{BF_Package}
Richard~D. Morey and Jeffrey~N. Rouder.
\newblock {BayesFactor}: Computation of {B}ayes factors for common designs, 2024.
\newblock R package version 0.9.12-4.7.

\bibitem{sinharay2005empirical}
Sandip Sinharay and Hal~S Stern.
\newblock An empirical comparison of methods for computing {B}ayes factors in generalized linear mixed models.
\newblock {\em Journal of Computational and Graphical Statistics}, 14(2):415--435, 2005.

\bibitem{jeffreys1935some}
Harold Jeffreys.
\newblock Some tests of significance, treated by the theory of probability.
\newblock {\em Mathematical Proceedings of the Cambridge Philosophical Society}, 31(2):203--222, 1935.

\bibitem{jeffreys1936further}
Harold Jeffreys.
\newblock Further significance tests.
\newblock {\em Mathematical Proceedings of the Cambridge Philosophical Society}, 32(3):416--445, 1936.

\bibitem{bartovs2023general}
Franti{\v{s}}ek Barto{\v{s}} and Eric-Jan Wagenmakers.
\newblock A general approximation to nested {B}ayes factors with informed priors.
\newblock {\em Stat}, 12(1):e600, 2023.

\bibitem{rostgaard2023simple}
Klaus Rostgaard.
\newblock Simple nested {B}ayesian hypothesis testing for meta-analysis, {C}ox, {P}oisson and logistic regression models.
\newblock {\em Scientific Reports}, 13(4731):1--11, 2023.

\bibitem{bickel2025small}
David~R Bickel.
\newblock A small-sample {B}ayesian information criterion that does not overstate the evidence, with an application to calibrating $p$-values from likelihood-ratio tests.
\newblock {\em Statistical Papers}, 66(3):1--17, 2025.

\bibitem{held2018p}
Leonhard Held and Manuela Ott.
\newblock On $p$-values and {B}ayes factors.
\newblock {\em Annual Review of Statistics and Its Application}, 5:393--419, 2018.

\bibitem{johnson2005bayes}
Valen~E Johnson.
\newblock Bayes factors based on test statistics.
\newblock {\em Journal of the Royal Statistical Society Series B: Statistical Methodology}, 67(5):689--701, 2005.

\bibitem{johnson2008properties}
Valen~E Johnson.
\newblock Properties of {B}ayes factors based on test statistics.
\newblock {\em Scandinavian Journal of Statistics}, 35(2):354--368, 2008.

\bibitem{yuan2008bayesian}
Ying Yuan and Valen~E Johnson.
\newblock Bayesian hypothesis tests using nonparametric statistics.
\newblock {\em Statistica Sinica}, 18(3):1185--1200, 2008.

\bibitem{wagenmakers2022approximate}
Eric-Jan Wagenmakers.
\newblock Approximate objective {B}ayes factors from $p$-values and sample size: The $3p\sqrt{n}$ rule.
\newblock {\em PsyArXiv}, 1:1--50, 2022.

\bibitem{neath2012bayesian}
Andrew~A Neath and Joseph~E Cavanaugh.
\newblock The {B}ayesian information criterion: Background, derivation, and applications.
\newblock {\em WIREs Computational Statistics}, 4(2):199--203, 2012.

\bibitem{mulder2022generalization}
Joris Mulder, Eric-Jan Wagenmakers, and Maarten Marsman.
\newblock A generalization of the {S}avage--{D}ickey density ratio for testing equality and order constrained hypotheses.
\newblock {\em The American Statistician}, 76(2):102--109, 2022.

\bibitem{masson2011tutorial}
Michael E~J Masson.
\newblock A tutorial on a practical {B}ayesian alternative to null-hypothesis significance testing.
\newblock {\em Behavior Research Methods}, 43:679--690, 2011.

\bibitem{nathoo2016bayesian}
Farouk~S Nathoo and Michael E~J Masson.
\newblock Bayesian alternatives to null-hypothesis significance testing for repeated-measures designs.
\newblock {\em Journal of Mathematical Psychology}, 72:144--157, 2016.

\bibitem{stan}
{Stan Development Team}.
\newblock {RStan}: the {R} interface to {Stan}, 2025.
\newblock R package version 2.32.7.

\bibitem{kooperberg1992logspline}
Charles Kooperberg and Charles~J Stone.
\newblock Logspline density estimation for censored data.
\newblock {\em Journal of Computational and Graphical Statistics}, 1(4):301--328, 1992.

\bibitem{gunel1974bayes}
Erdogan Gunel and James Dickey.
\newblock Bayes factors for independence in contingency tables.
\newblock {\em Biometrika}, 61(3):545--557, 1974.

\bibitem{Begley2012-ur}
C~Glenn Begley and Lee~M Ellis.
\newblock Raise standards for preclinical cancer research.
\newblock {\em Nature}, 483(7391):531--533, 2012.

\bibitem{Hwang2016-pw}
Thomas~J Hwang, Daniel Carpenter, Julie~C Lauffenburger, Bo~Wang, Jessica~M Franklin, and Aaron~S Kesselheim.
\newblock Failure of investigational drugs in late-stage clinical development and publication of trial results.
\newblock {\em JAMA Internal Medicine}, 176(12):1826--1833, 2016.

\bibitem{Sun2022-ma}
Duxin Sun, Wei Gao, Hongxiang Hu, and Simon Zhou.
\newblock Why 90\% of clinical drug development fails and how to improve it?
\newblock {\em Acta Pharmaceutica Sinica B}, 12(7):3049--3062, 2022.

\bibitem{Miron2020-gm}
Laura Miron, Rafael~S Gon{\c{c}}alves, and Mark~A Musen.
\newblock Obstacles to the reuse of study metadata in {ClinicalTrials.gov}.
\newblock {\em Scientific Data}, 7(443):1--14, 2020.

\bibitem{lee2014bayesian}
Michael~D Lee and Eric-Jan Wagenmakers.
\newblock {\em Bayesian cognitive modeling: A practical course}.
\newblock Cambridge University Press, 2014.

\end{thebibliography}






\clearpage

\section*{Appendix A: Preprocessing Steps}

\paragraph{CTG Download.}
We downloaded the CTG results for completed studies with posted results as a JSON export (all available fields) from \url{https://clinicaltrials.gov/search?aggFilters=results:with,status:com} on August 7th, 2025. During data cleaning, we excluded four records with sample-size errors identified while reviewing candidate type I errors at $\alpha=.01$.

\paragraph{Record Construction.}
We transformed each ClinicalTrials.gov JSON record into flat, analysis-level rows prior to $p$-value cleaning and method harmonization. For each study (\texttt{nctId}), we iterated over outcome measures with at least one analysis and, for every outcome–analysis pair, emitted one row per denominator count with units “Participants” and with \texttt{groupId} in the analysis’s declared \texttt{groupIds}. We carried forward study metadata (e.g., \texttt{briefTitle}, \texttt{studyType}), design-information fields (\texttt{interventionModel}, \texttt{observationalModel}, \texttt{timePerspective}, \texttt{allocation}), one-hot encoded phases (\texttt{EARLY\_PHASE1}, \texttt{PHASE1}, \texttt{PHASE2}, \texttt{PHASE3}, \texttt{PHASE4}), and MeSH terms/IDs for conditions and interventions from the \texttt{derivedSection}. We also flagged whether the study included any references of type \texttt{RESULTS} or \texttt{DERIVED}.

\paragraph{Analysis Identifier.}
Each row was keyed by a stable, hash-based identifier,
\[
\texttt{analysisId} \;=\; \operatorname{MD5}\!\big(\texttt{nctId} + \mathrm{JSON}(\texttt{outcome};\ \text{sorted keys}) + \mathrm{JSON}(\texttt{analysis};\ \text{sorted keys})\big),
\]
which yields reproducible IDs for unique outcome–analysis pairs within a study (independent of JSON key order). This \texttt{analysisId} serves as the unit of aggregation in downstream preprocessing (e.g., computing sample sizes and $\mathit{eJAB}$).

\paragraph{Schema.}
Fields include: \texttt{analysisId}, \texttt{nctId}, \texttt{briefTitle}, \texttt{outcomeType}, \texttt{outcomeTitle}, \texttt{groupDescription}, \texttt{units}, \texttt{groupId}, \texttt{value}, \texttt{pValue}, \texttt{statisticalMethod}; MeSH fields (\texttt{condition\_mesh}, \texttt{condition\_ids}, \texttt{intervention\_mesh}, \texttt{intervention\_ids}); \texttt{hasResultsOrDerived}; study-design fields; and phase indicators. The field \texttt{value} records the number of participants for the given outcome group/arm.

\paragraph{Standardization of Reported $p$-Values.}
Reported $p$-values often included textual qualifiers (e.g., ``\textless 0.05''). We standardized them as follows:
\begin{enumerate}
  \item Flag the presence of a ``\textless{}'' qualifier: \texttt{has\_less\_than \(\gets\) grepl("<", pValue)}.
  \item Extract the numeric component via regex: \(\texttt{pValue} \gets \texttt{str\_extract(pValue, "[0-9]*\\.?[0-9]+")}\).
  \item Drop rows with missing or blank numeric components.
  \item Coerce \texttt{pValue} to numeric and retain only valid probabilities: \(0 < \texttt{pValue} < 1\).
  \item Exclude rows with zero sample size, which are incompatible with downstream model mappings.
\end{enumerate}

\paragraph{Harmonization of Statistical Method Names.}
Free-text \texttt{statisticalMethod} labels were mapped to canonical test families using regular-expression rules (\texttt{str\_detect}). Observations with no match were excluded. The exact mapping rules are:
\begin{lstlisting}[language=R,caption={Regex-based mapping of free-text methods to canonical families}]
statisticalMethod <- case_when(
  # --- T-TESTS ---
  str_detect(originalMethod, regex("\\bt[- ]?test", ignore_case = TRUE)) &
    str_detect(originalMethod, regex("one[- ]?sample", ignore_case = TRUE)) ~ "1-ttest",

  str_detect(originalMethod, regex("\\bt[- ]?test", ignore_case = TRUE)) &
    str_detect(originalMethod, regex("paired", ignore_case = TRUE)) ~ "1-ttest",

  str_detect(originalMethod, regex("\\bt[- ]?test", ignore_case = TRUE)) ~ "2-ttest",

  # --- CONDITIONAL LOGISTIC REGRESSION ---
  str_detect(originalMethod, regex("logistic", ignore_case = TRUE)) &
    str_detect(originalMethod, regex("regression", ignore_case = TRUE)) &
    str_detect(originalMethod, regex("conditional", ignore_case = TRUE)) ~ "clogit",

  # --- LOGISTIC REGRESSION ---
  str_detect(originalMethod, regex("logistic", ignore_case = TRUE)) &
    str_detect(originalMethod, regex("regression", ignore_case = TRUE)) ~ "glm",

  # --- COX ---
  str_detect(originalMethod, regex("cox", ignore_case = TRUE)) &
    !str_detect(originalMethod, regex("wilcoxon|mantel|signed|rank", ignore_case = TRUE)) ~ "cox",

  # --- LOGRANK / MANTEL-COX ---
  str_detect(originalMethod, regex("log ?rank|mantel[- ]?cox", ignore_case = TRUE)) &
    !str_detect(originalMethod, regex("wilcoxon|signed", ignore_case = TRUE)) ~ "logrank",

  # --- MANN-WHITNEY ---
  str_detect(originalMethod, regex("mann[- ]?whitney", ignore_case = TRUE)) |
    (str_detect(originalMethod, regex("wilcoxon", ignore_case = TRUE)) &
       str_detect(originalMethod, regex("rank", ignore_case = TRUE)) &
       !str_detect(originalMethod, regex("signed", ignore_case = TRUE))) ~ "Utest",

  # --- WILCOXON SIGNED-RANK ---
  str_detect(originalMethod, regex("wilcoxon", ignore_case = TRUE)) &
    str_detect(originalMethod, regex("signed", ignore_case = TRUE)) &
    str_detect(originalMethod, regex("rank", ignore_case = TRUE)) ~ "wilcox",

  # --- KRUSKAL-WALLIS ---
  str_detect(originalMethod, regex("kruskal|wallis", ignore_case = TRUE)) ~ "Htest",

  # --- CHI-SQUARE ---
  str_detect(originalMethod, regex("chi[- ]?square|χ2|chi2|chisq", ignore_case = TRUE)) ~ "chisq",

  # --- REPEATED MEASURES ---
  str_detect(originalMethod, regex("repeated[- ]?measures?", ignore_case = TRUE)) ~ "rANOVA",

  # --- LINEAR REGRESSION ---
  str_detect(originalMethod, regex("linear", ignore_case = TRUE)) &
    str_detect(originalMethod, regex("regression", ignore_case = TRUE)) ~ "lm",

  # --- ANOVA ---
  str_detect(originalMethod, regex("anova|analysis of variance", ignore_case = TRUE)) ~ "ANOVA",

  TRUE ~ NA_character_
)
\end{lstlisting}

\paragraph{Adjustment for One-Sided $p$-Values.}
When free text indicated a one-sided analysis (patterns: \texttt{one-sided}, \texttt{1-sided}), we converted to a two-sided $p$-value by doubling and truncating at 1:
\[
\texttt{pValue} \leftarrow \min\{2 \cdot \texttt{pValue},\ 1\}.
\]
This rule was applied only when \texttt{str\_detect} flagged these patterns in \texttt{originalMethod}.

\paragraph{analysisId Sample Size QC.}
Certain tests should yield exactly one row per \texttt{analysisId} (e.g., one-sample $t$-test, Wilcoxon signed-rank test). To enforce consistency, we dropped analyses in which such tests appeared with multiple rows under the same \texttt{analysisId}.

\paragraph{Aggregation to the Test Level and Parameter Construction.}
We aggregated records at the \texttt{analysisId} (test) level. Fields invariant within a test were taken from the first row—e.g., \texttt{nctId}, \texttt{pValue}, \texttt{statisticalMethod}, \texttt{studyType}, phase indicators (\texttt{EARLY\_PHASE1}, \texttt{PHASE1}, \texttt{PHASE2}, \texttt{PHASE3}, \texttt{PHASE4}), \texttt{outcomeTitle}, \texttt{outcomeType}, \texttt{groupDescription}, \texttt{conditionIds}, \texttt{briefTitle}, and \texttt{has\_less\_than}.

We defined the effective sample size $n$ per test as:
\[
n =
\begin{cases}
\ \texttt{first(value)}, & \text{one-sample $t$-test or Wilcoxon signed-rank};\\[2pt]
\ \max \texttt{value}, & \text{conditional logistic regression};\\[2pt]
\ 0.5 \sum \texttt{value}, & \text{Cox PH or log-rank};\\[2pt]
\ \sum \texttt{value}, & \text{otherwise.}
\end{cases}
\]
We also set method-specific parameters required by $\mathit{eJAB}$:
\[
\begin{aligned}
    &\texttt{I} =
\begin{cases}
\ n_{\text{rows within test}}, & \text{ANOVA, Kruskal–Wallis, repeated measures};\\
\ \text{NA}, & \text{otherwise},
\end{cases} \\
    &(\texttt{R}, \texttt{C}) =
\begin{cases}
\ (n_{\text{rows within test}},\ 2), & \text{chi-squared test};\\
\ (\text{NA},\ \text{NA}), & \text{otherwise}.
\end{cases}
\end{aligned}
\]

\paragraph{Mapping to JAB Models and Computation of eJAB.}
We mapped canonical \texttt{statisticalMethod} labels to the \texttt{JAB01} model strings required for $\mathit{eJAB}_{01}$. Mapping was performed only for tests included in our analysis (i.e., excluding chi-squared, Cox PH, log-rank, and repeated-measures designs):
\[
\begin{aligned}
&\text{Two-sample $t$-test, paired $t$-test, one-sample $t$-test} \rightarrow \texttt{"t-test"},\\
&\text{Linear regression} \rightarrow \texttt{"linear\_regression"},\\
&\text{Logistic regression, conditional logistic regression} \rightarrow \texttt{"logistic\_regression"},\\
&\text{ANOVA} \rightarrow \texttt{"anova"},\quad
\text{Kruskal–Wallis} \rightarrow \texttt{"kruskal\_wallis"},\\
&\text{Mann–Whitney} \rightarrow \texttt{"mann\_whitney"},\quad
\text{Wilcoxon} \rightarrow \texttt{"wilcoxon"}.
\end{aligned}
\]
Given \((n, \texttt{pValue}, \texttt{model}, R, C, I)\) for each test, we computed $\mathit{eJAB}_{01}$ via
\[
\texttt{JAB} \;=\; \texttt{JAB01}(n,\ \texttt{pValue},\ \texttt{model};\ R,\ C,\ I).
\]

\paragraph{Exclusions and Final Dataset.}
We excluded rows with missing \texttt{model}, missing \texttt{JAB} or \texttt{pValue}, and tests with $n=1$. The final dataset, \texttt{study3\_summary}, contains one record per \texttt{analysisId} with standardized \texttt{pValue}, computed \texttt{JAB}, mapped method family, phase indicators, and associated metadata.


\section*{Appendix B: Supplementary Figures}

\clearpage

\begin{figure*}[!ht]
    \centering
    \includegraphics[height=0.95\textheight]{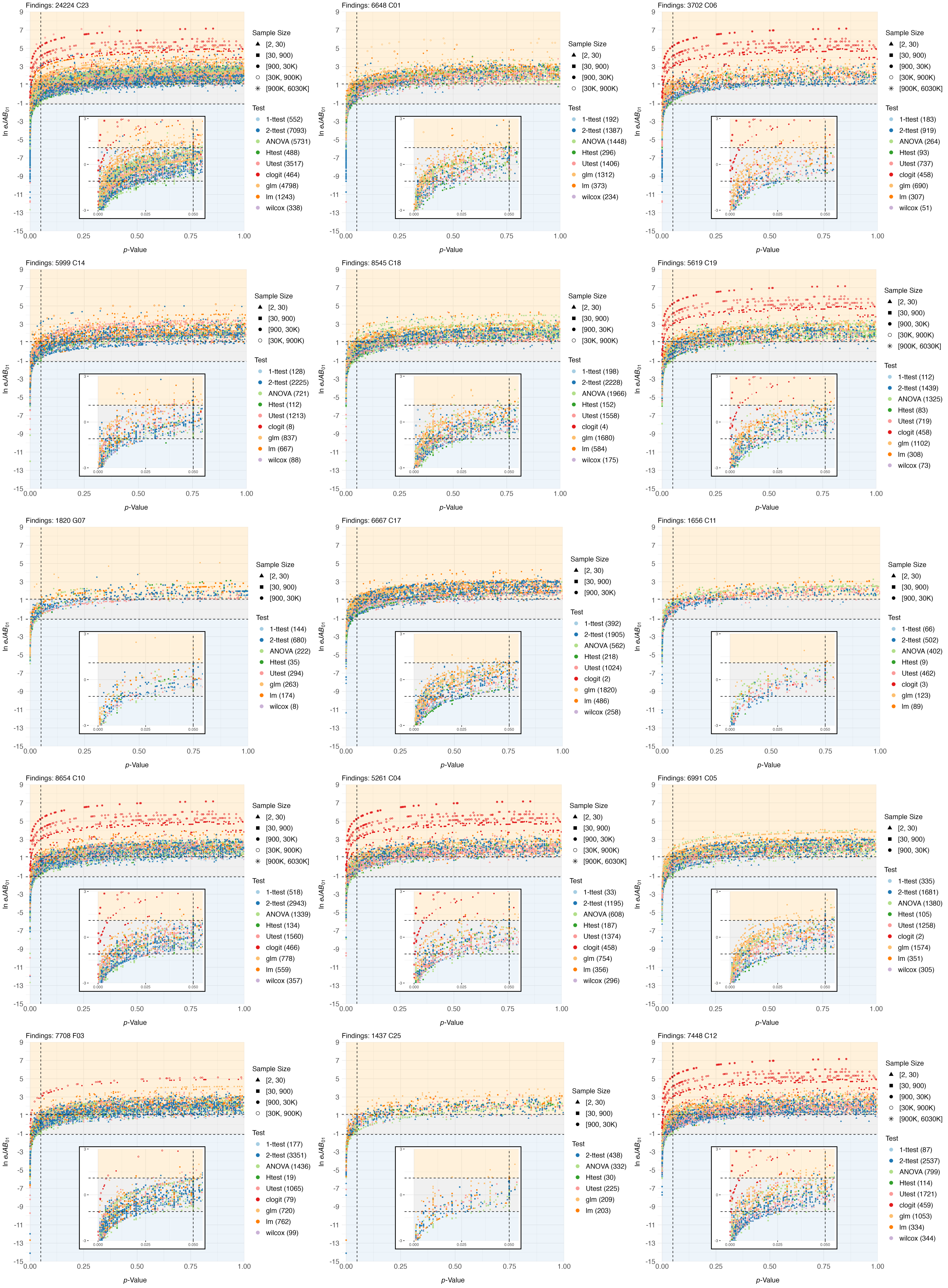}
    \caption{Scatterplots of $\ln \mathit{eJAB}_{01}$ versus $p$-value by MeSH condition (part 1 of 3), with inset zooms for $p\leqslant.05$. Shaded bands denote evidence categories.}
    \label{fig8}
\end{figure*}


\clearpage

\begin{figure*}[!ht]
    \centering
    \includegraphics[height=0.95\textheight]{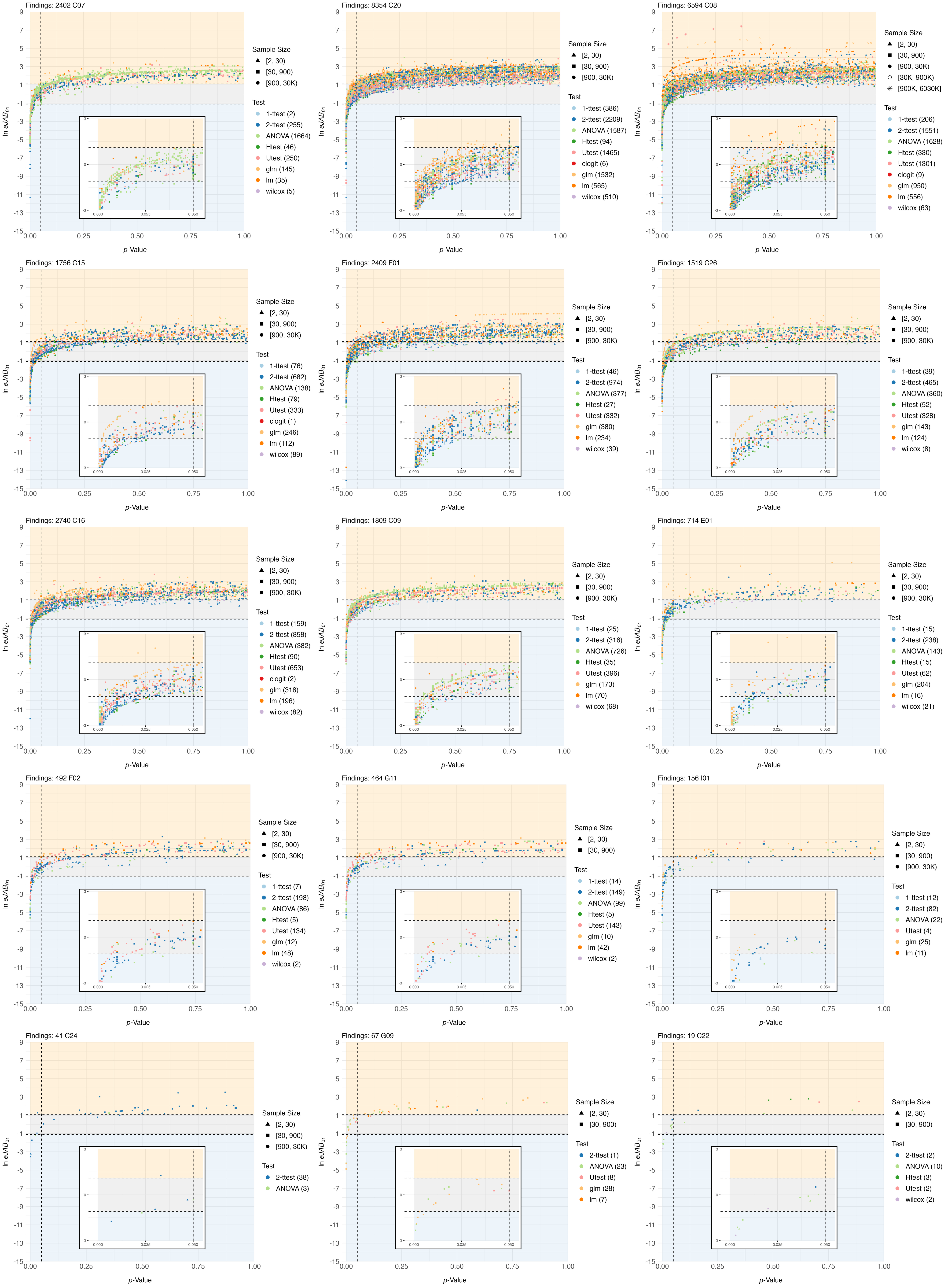}
    \caption{Scatterplots of $\ln \mathit{eJAB}_{01}$ versus $p$-value by MeSH condition (part 2 of 3), with inset zooms for $p\leqslant.05$. Shaded bands denote evidence categories.}
    \label{fig9}
\end{figure*}


\clearpage

\begin{figure*}[!htb]
    \centering
    \includegraphics[width=0.95\textwidth]{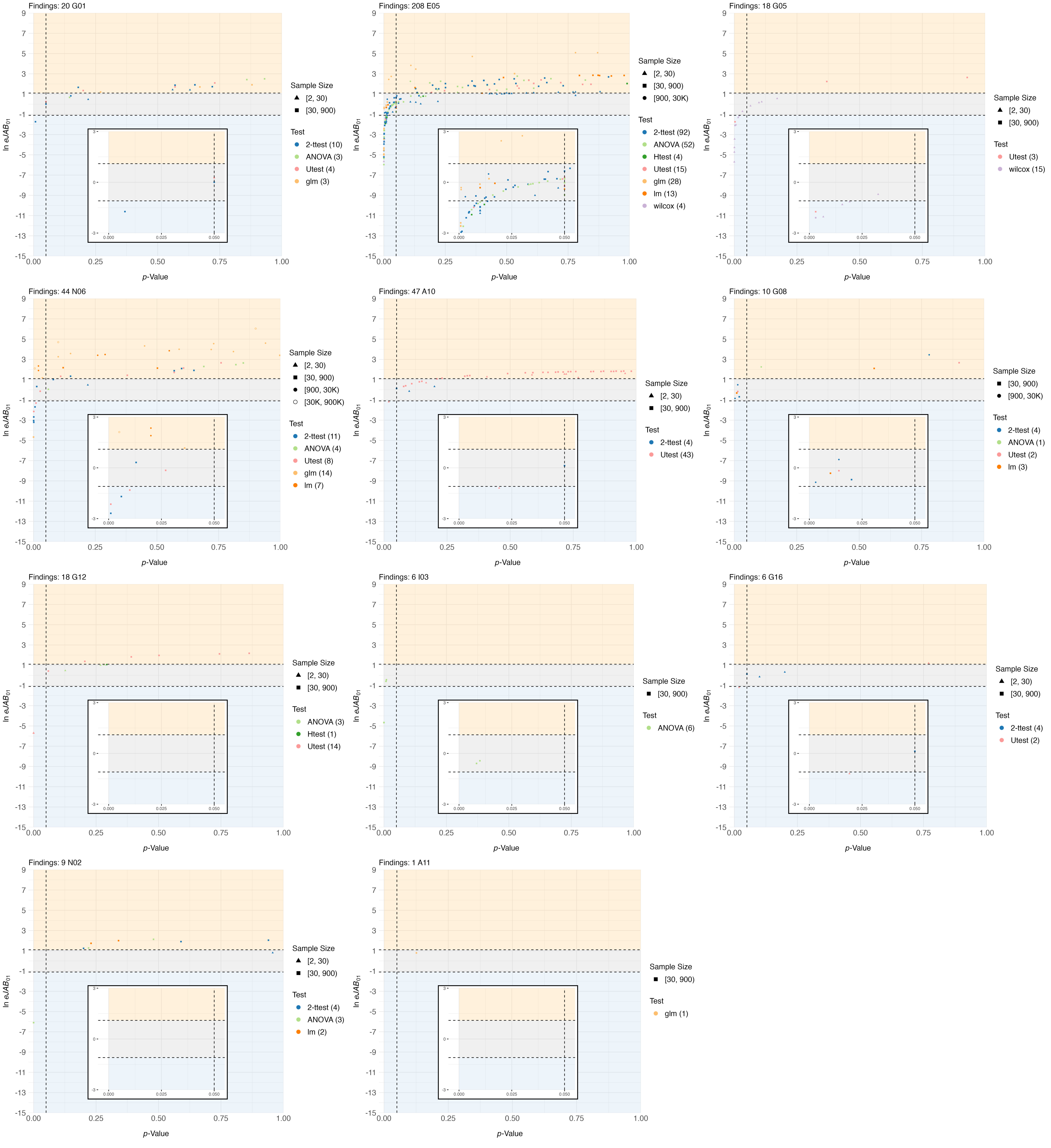}
    \caption{Scatterplots of $\ln \mathit{eJAB}_{01}$ versus $p$-value by MeSH condition (part 3 of 3), with inset zooms for $p\leqslant.05$. Shaded bands denote evidence categories.}
    \label{fig10}
\end{figure*}


\clearpage

\begin{figure*}[!htb]
    \centering
    \includegraphics[width=0.8\textwidth]{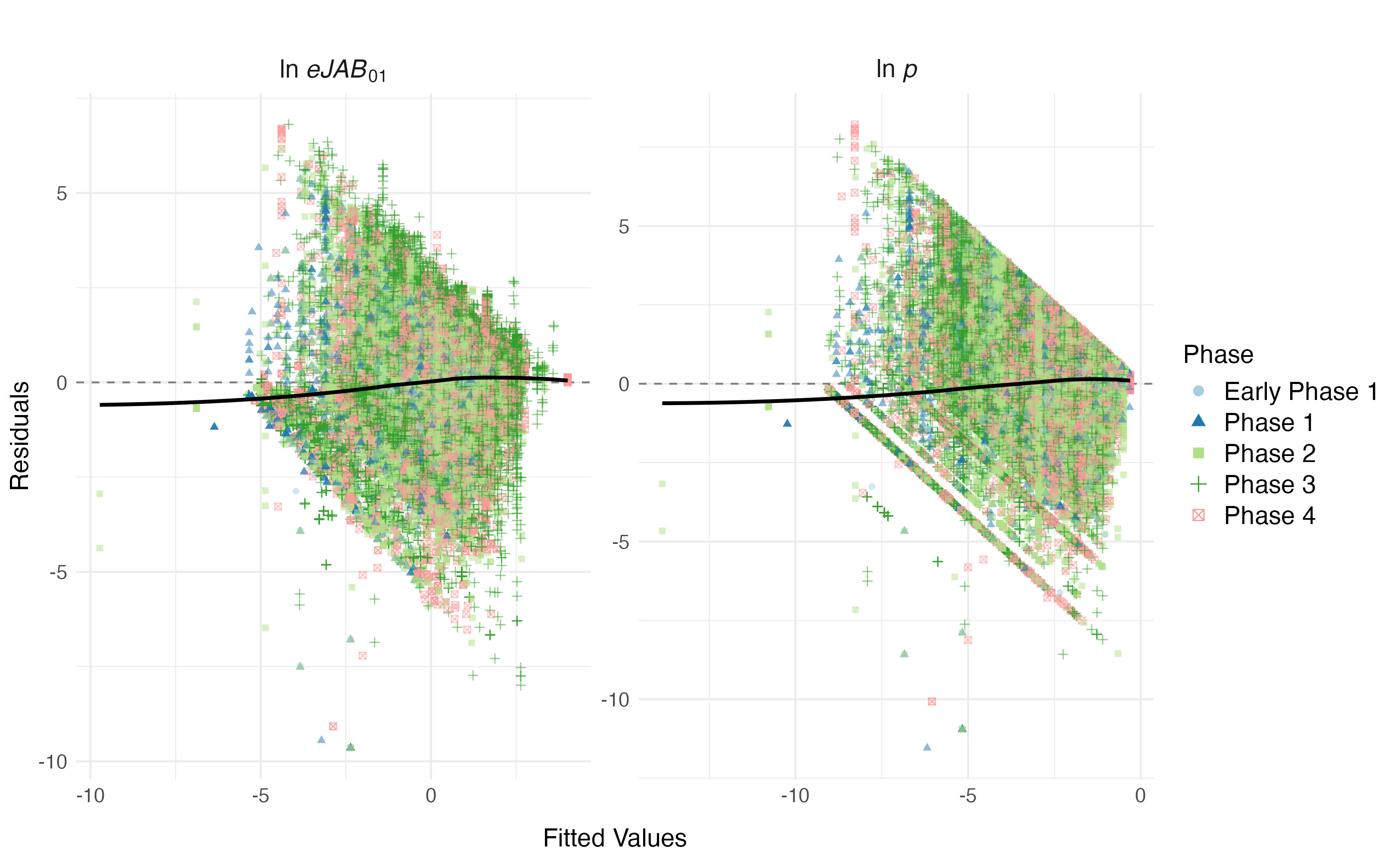}
    \caption{Residuals versus fitted values from random-intercept models for $\ln\mathit{eJAB_{01}}$ (left) and $\ln{p}$ (right). The dashed horizontal line marks zero residuals and the solid curve is a LOESS smooth. Nearly parallel slanted bands reflect heaping at common reported $p$ thresholds.}
    \label{fig11}
\end{figure*}

\begin{figure*}[!htb]
    \centering
    \includegraphics[width=0.8\textwidth]{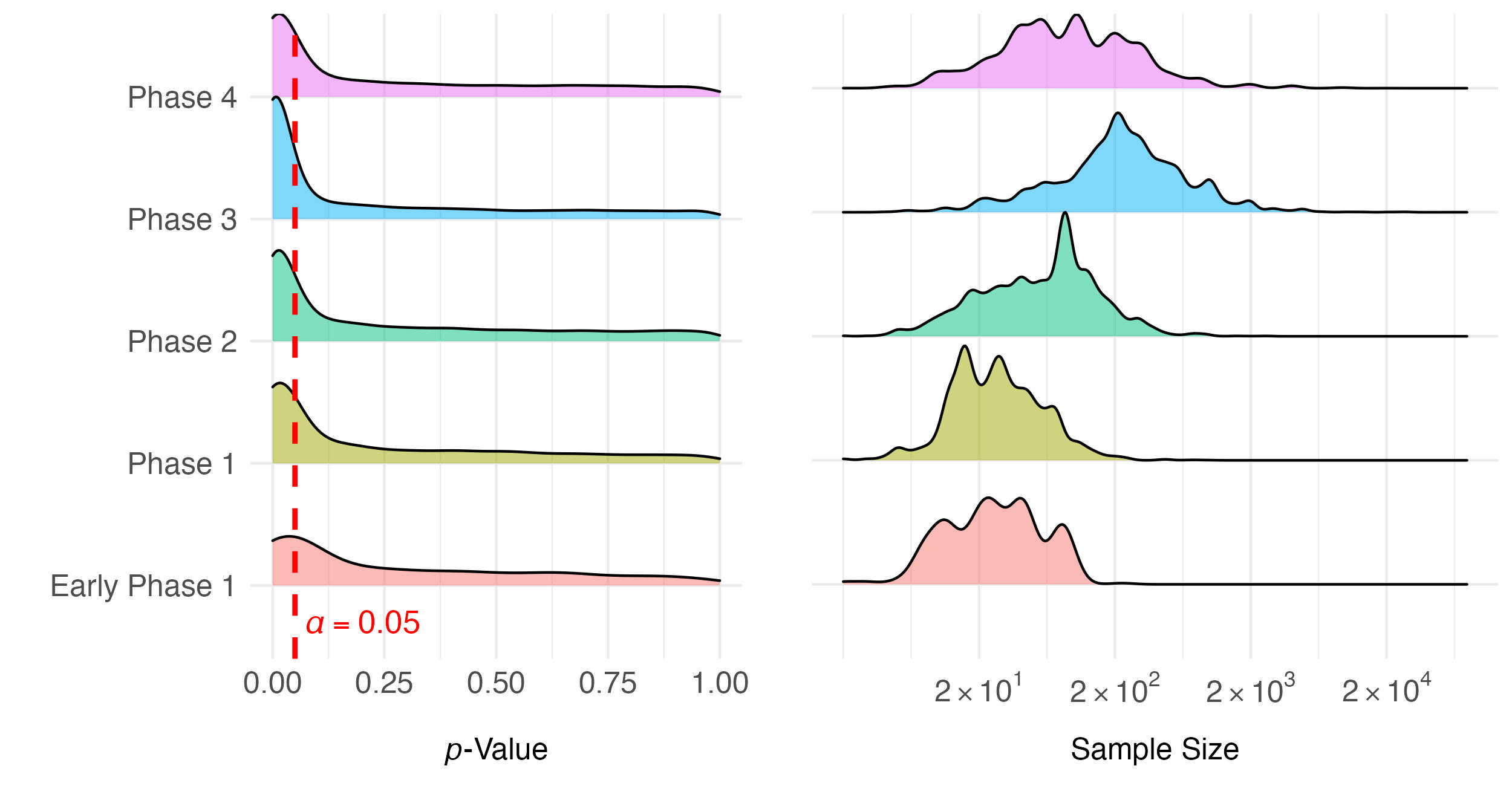}
    \caption{Distributions of $p$-values (left; vertical dashed line at~.05) and sample size (right) by clinical trial phase.}
    \label{fig12}
\end{figure*}

\end{document}